# HyPPI NoC: Bringing Hybrid Plasmonics to an Opto-Electronic Network-on-Chip

Vikram K. Narayana, Shuai Sun, Armin Mehrabian, Volker J. Sorger, and Tarek El-Ghazawi
*Department of Electrical and Computer Engineering*
*The George Washington University*
*Washington, D.C. 20052*
{*vikram,sunshuai,armin,sorger,tarek*}@gwu.edu

*Abstract*—As we move towards an era of hundreds of cores, the research community has witnessed the emergence of opto-electronic network on-chip designs based on nanophotonics, in order to achieve higher network throughput, lower latencies, and lower dynamic power. However, traditional nanophotonics options face limitations such as large device footprints compared with electronics, higher static power due to continuous laser operation, and an upper limit on achievable data rates due to large device capacitances. Nanoplasmonics is an emerging technology that has the potential for providing transformative gains on multiple metrics due to its potential to increase the light-matter interaction. In this paper, we propose and analyze a hybrid opto-electric NoC that incorporates Hybrid Plasmonics Photonics Interconnect (HyPPI), an optical interconnect that combines photonics with plasmonics. We explore various opto-electronic network hybridization options by augmenting a mesh network with HyPPI links, and compare them with the equivalent options afforded by conventional nanophotonics as well as pure electronics. Our design space exploration indicates that augmenting an electronic NoC with HyPPI gives a performance to cost ratio improvement of up to 1.8×. To further validate our estimates, we conduct trace based simulations using the NAS Parallel Benchmark suite. These benchmarks show latency improvements up to 1.64×, with negligible energy increase. We then further carry out performance and cost projections for fully optical NoCs, using HyPPI as well as conventional nanophotonics. These futuristic projections indicate that all-HyPPI NoCs would be two orders more energy efficient than electronics, and two orders more area efficient than all-photonic NoCs.

*Keywords*-Hybrid Plasmon-Photonics; Network-on-Chip; Hybrid NoC;

## I. Introduction and Motivation

With the drive towards an increasing number of cores over the past decade, the field has witnessed network-on-chip (NoC) architectures becoming mainstream in high-performance computing and embedded processors. However, challenges arising from the steadily rising memory demands of applications, as well as the rising power budget for on-chip networks, has led researchers to actively investigate nanophotonics-based opto-electronic networks [1].

Photonic interconnects have been considered as a promising on-chip option in the ITRS roadmap since the mid-2000s. Because of the parallelism of bosons, it is able to support Wavelength Division Multiplexing (WDM) for a higher bandwidth compared to electrical interconnects. Furthermore, the low optical attenuation of photonics enables on-chip communications with small energy losses during light propagation. Photonics is thus efficient for passive operations such as data transmission. Examples of proposed photonic NoCs include: Corona [2], a NoC with multi-write single-read optical loops accessed through token-based arbitration; LumiNoC, a NoC with multiple optical subnets [3]; Flexishare [4], a multi-stage optical crossbar interconnect; and ATAC, a hybrid NoC that uses an optical loop for broadcast operations [5].

While nanophotonics has proven to be promising, active operations that involve light manipulation face some fundamental limitations. As such, purely photonic devices are diffraction limited, which means that device sizes below the wavelength scale will lead to a leakage of light into the surrounding medium. This often leads to bulky devices that are an order of magnitude larger than electronics. For example, consider a typical nanophotonic link shown in Fig. 1a. Microring resonators (MRRs) tuned to the required wavelength are utilized for modulating the laser by selectively diverting light from the main waveguide to transmit binary data [6]. The ability to rapidly tune and detune from the default wavelength requires MRRs to have a high quality factor (Q factor), which affects the photon lifetime and device scaling. This usually leads to an upper limit on operating frequency and larger device footprint. As seen in Fig. 1a, the MRR diameter can be as small as 5 $\mu$m but requires an additional 15 $\mu$m spacing to eliminate thermal crosstalk [7], leading to a single MRR area that is comparable to the size of a 5×5 crossbar for 64-bit flit size*. Consequently, challenges arise in dense integration with nanometer-scale electronics. Furthermore, there are higher power demands due to thermal trimming required for the MRRs [1].

To overcome these limitations, we have recently proposed Hybrid Plasmonic Photonic Interconnect (HyPPI), an interconnect that relies on combining nanophotonics with an emerging technology called plasmonics [9]. While HyPPI has been recently studied earlier, its implications at the on-chip network level has not been explored, and is thus the focus of the investigations in this paper.

---

*Router data obtained from DSENT [8] for 11 nm tech. node.

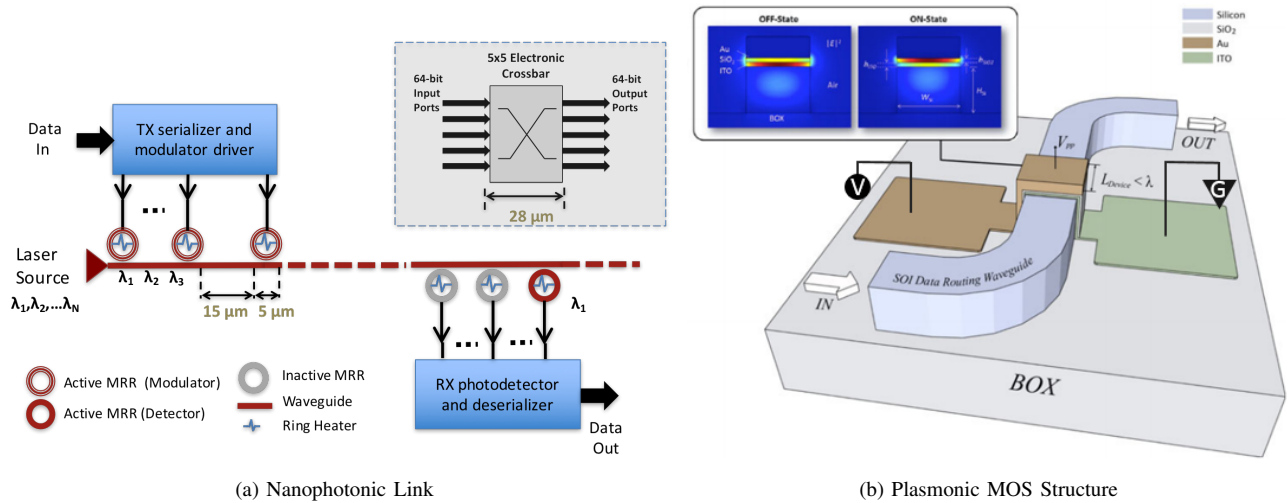

(a) Nanophotonic Link  (b) Plasmonic MOS Structure

Figure 1. Nanophotonics and Plasmonics.

## II. Hybrid Plasmon Photon Interconnects (HyPPI)

As discussed, conventional nanophotonics is diffraction limited. To overcome this hurdle, plasmonics and metal optics have emerged as an alternative. Plasmonics essentially comprises of light waves that propagate along the surface of a conductor, and involves coupled electron and light oscillations propagating on a metal waveguide. Surface plasmons, while having the same frequency as the incident light, have much smaller wavelengths, allowing very small footprint devices. This in turn enables ultrafast operating frequency due to much lower device capacitance. However, plasmonics suffers from high ohmic losses, restricting propagation distances to only a few microns.

As described, both photonic and plasmonic interconnects have their own shortages in either individual device performance or propagation distance. To overcome these, photonic interconnects can be combined with plasmonic interconnects to achieve higher operating frequency, higher energy efficiency and longer propagation distance with smaller on-chip footprint (HyPPI) [9]. A modulator based on a plasmonic metal oxide semiconductor (MOS) type structure [10] is shown Fig. 1b. Unlike several tens of microns needed by MRR-based nanophotonics, this device has a size of the order of $1\mu m$. By using a conventional photonic SOI waveguide, the propagation distance of HyPPI can be extended to the centimeter range while maintaining ultrafast speed and low energy consumption due to the plasmonic modulator.

## III. Building Hybrid On-Chip Networks

HyPPI is an excellent candidate as a point-to-point link, to replace electronic links in a network-on-chip (NoC). However, due to reliance on the electronic routers for directing flits across the NoC, there are a lot of optical to electrical (O-E) and electrical to optical (E-O) conversions that occur as a result. For instance, consider the Mesh NoC shown in Fig. 2a. Each one of the 'Regular Link' can be optical, however, a node communicating from the left end to the right end will incur several O-E-O conversions. One possible approach to address this issue is through the use of express links [11]. An example with 2 hops express links in the horizontal direction is shown in Fig. 2b. Since additional links demand a larger number of ports from the participating routers, we consider express links only in the horizontal direction.

The other option is to use an all-optical NoC, see Figure 2c. However, in our opinion, completely optical NoCs are not yet fully mature for migration from contemporary electronic networks. For instance, earlier work showed that links in fully-optical NoCs are underutilized in real applications due to their low injection rates [12]. We thus believe that it is better to deploy photonic links only for long-range traffic and for nodes that communicate heavily. Furthermore, with the lack of memory storage in optics (no flip flops or registers or buffers), an all-optical network will require a suitable infrastructure for arbitration and/or routing, with proposed approaches using token-based arbitration [2] or a parallel electronic path for channel setup [13]. Thus, we prefer to adopt the cheaper and well-understood and easily routable electronics for short distances. Furthermore, due to additional clock cycles overhead in opto-electric conversions, optical links become inferior for short distance traffic between, for instance, neighboring core routers.

We thus believe that as a first step, augmenting electronic NoCs with long optical links (nanophotonics or HyPPI) is a viable path forward, Fig. 2a, and Fig. 2b. With Electronic, Photonic, Plasmonic, and HyPPI as our options, we need to choose suitable combinations in building the NoCs depicted. As such, performance metrics are required in order to carry out a design-space exploration. Several metrics have been utilized in the past, such as Latency, Throughput, Power/Energy, and Area, and combinations thereof such as performance per watt or energy delay product. In this work, we define a unified figure of merit called CLEAR, described as follows. CLEAR is a higher-is-better metric.

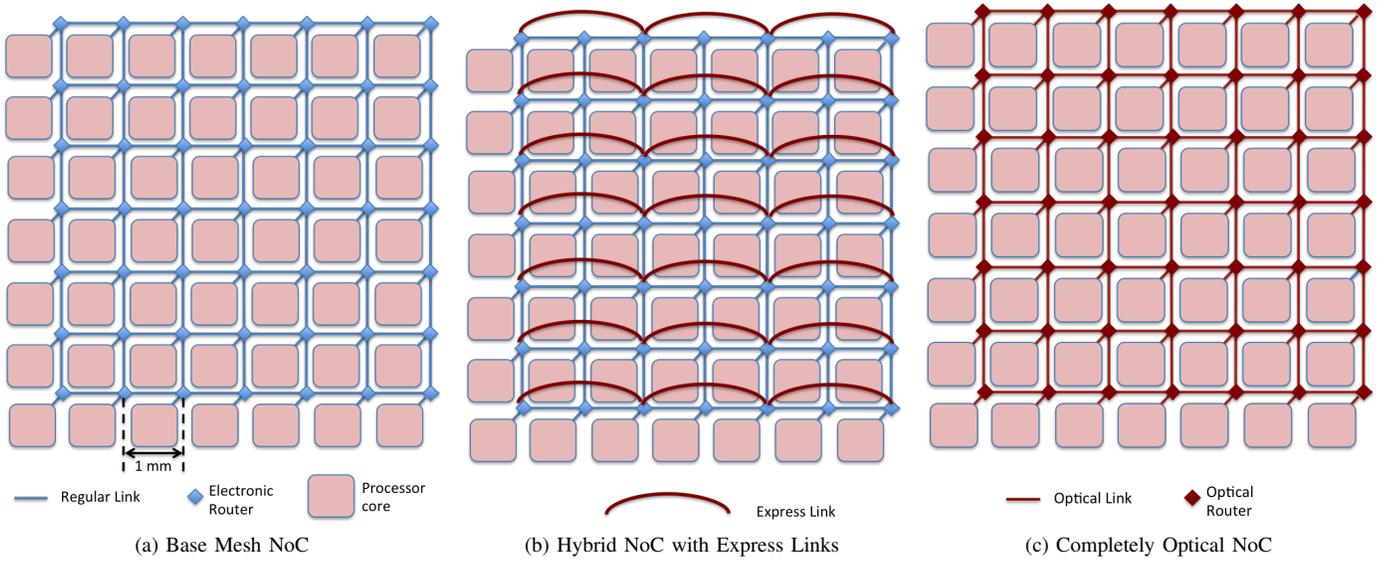

Figure 2. Networks evaluated for different technology options. The small number of cores is for illustration only. Express Links shown are for Hops = 2. All links are bidirectional (not shown).

### A. Evaluating Individual Links

The Capability to Latency-Energy-Area Ratio (CLEAR) is a unified figure of merit, equation (1). It can be regarded as a performance to cost ratio, where the performance capability (C) is the data capacity (in Gbps) of a link. For the cost part, it includes the point-to-point latency (L, in *ps*) of the link, the overall energy efficiency (E, in fJ/bit) and the on-chip area (A, in $\mu m^2$). These three factors in the denominator denote the time, energy and space cost to achieve a certain Capability. And therefore, higher CLEAR value demonstrates a link with faster data rate and longer propagation distance, while its latency, energy consumption and on-chip footprint are relatively smaller.

$$CLEAR\ (link) = \frac{Capability}{Latency\ \times\ Energy\ \times\ Area} \quad (1)$$

Since this figure of merit is used for the performance comparison among different interconnect options, there is no need to use SI units for all the factors in CLEAR if we only need to know the relative values.

We now apply CLEAR to the different link technology options, Electronic, Photonic, Plasmonic, and HyPPI. The interconnect parameters for calculating CLEAR are based

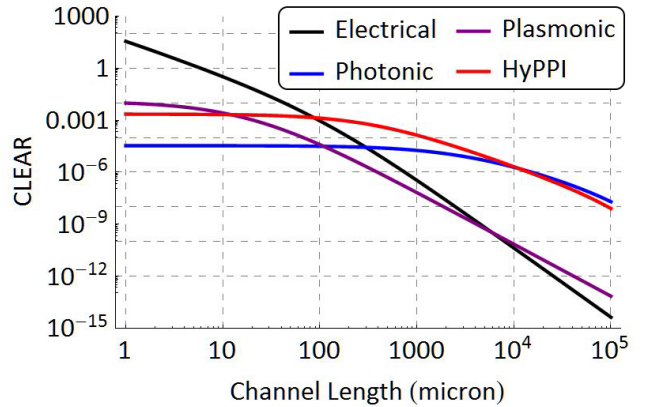

Figure 3. CLEAR figure of merit for links.

on those in the literature [14], [9], and are summarized in Table I. The electrical link parameters are borrowed from the 14 nm technology node ITRS roadmap. A plot of CLEAR is shown in Fig. 3. For this evaluation, all links are used in point-to-point mode. For photonics, ring modulators and ring detectors are utilized, leading larger area overheads as discussed in the introduction. Electronics is best suited for short interconnects, both logic level and intra-processor communication. This is due to the lower static power, as well as lower area requirements. For larger lengths, such as inter-core distances, HyPPI is more favorable, as the higher static power is amortized over the distance, while electronics begins to incur higher power due to repeaters. Pure plasmonic is suited only for very tiny distances (few microns) due to huge ohmic losses. Thus, pure plasmonics is

Table I
PHOTONIC, PLASMONIC, HYPPI LINK PARAMETERS

| Parameters | Photonic | Plasmonic | HyPPI |
|---|---|---|---|
| **Laser** | | | |
| Efficiency (%) | 25 | 20 | 20 |
| Area ($\mu m^2$) | 200 | 0.003 | 0.003 |
| **Modulator** | | | |
| Speed (Gbps) | 25 (25) | 59[†](50) | 2100[†](50) |
| Energy Efficiency (fJ/bit) | 2.77* | 6.8* | 4.25* |
| Insertion Loss (dB) | 1.02* | 1.1 | 0.6 |
| Extinction Ratio (dB) | 6.18* | 17 | 12 |
| Area ($\mu m^2$) | 100 | 4 | 1 |
| Capacitance (fF) | 16 | 14 | 0.94 |
| Bias Voltage (V) | -2.2∼0.4 | 0.7 | 2∼3 |
| **Photodetector** | | | |
| Speed (Gb/s) | 40/40 | 50/700 | 50/700 |
| Energy Efficiency (fJ/bit) | 0* | 0.14* | 0.14* |
| Responsivity (A/W) | 0.8 | 0.1 | 0.1 |
| Area ($\mu m^2$) | 100 | 4 | 4 |
| **Waveguide** | | | |
| Propagation Loss (dB/cm) | 1 | 440 | 1 |
| Coupling Loss (dB) | - | 0.63 | 1 |
| Pitch ($\mu m$) | 4 | 0.5 | 1 |
| Width ($\mu m$) | 0.35 | 0.1 | 0.35 |

* These values are used only for a bare link-level comparison; at the NoC system level, we instead use a modified version of the DSENT tool to compute the optimized values, which incorporates associated circuits for drivers and SERDES.

[†] These data rates are supported by the links and adopted for our bare links comparison; however, at the system level, the SERDES circuitry poses an upper limit on the data rate, as indicated in the parenthesized values used in our NoC system simulations.

not considered any further in our network level explorations. Photonics becomes suitable for lengths beyond 20 mm, which explains their popularity in inter-node interconnects in supercomputers.

Note, however, that these link level evaluations may not be suitable directly at the network level. For instance, latency measured as a continuous analog value no longer holds for NoCs, where short and medium length latencies are accommodated within a clock cycle. Also, the photonic link evaluated here ignores the large power requirements for thermal trimming of rings required in NoCs (see Section I), thus the energy per bit would be larger than in Table I. Furthermore, our link-level evaluations assumed the data rates (modulator speeds) listed in Table I, which gives the peak device capability. For instance, it is 2.10 Tb/s for HyPPI. In contemporary technology, however, we are limited to the capability of the driver and SERDES electronics circuitry, which we found to be capable of 50 Gb/s based on our experiments with DSENT [8]. These network level implications are further captured in our CLEAR evaluations at the NoC level.

*B. Hybrid Networks Exploration*

As discussed, we use CLEAR for the design space exploration of the different technology options for the NoCs shown in Fig. 2a and Fig. 2b. For the system level, we adjust the individual factors and also add a factor R. Thus, the expression for CLEAR is,

$$\frac{\left(\sum_{i=1}^{all\ links} C_i\right)/N}{Latency\ (clks)\ \times\ Power\ \times\ Area\ \times\ R} \quad (2)$$

Where $C_i$ is the capacity of link $i$, and $N$ is the number of nodes. The numerator is the aggregate link capacities averaged across all nodes. A larger numerator means potentially more paths for packets to traverse. The factor $R$ is the rate of increase of utilization,

$$R = \frac{dU}{dr} \quad (3)$$

where $U$ is the average link utilization of the network, and $r$ is the traffic injection rate. If $R$ is large, then as the injection rate is increased, link utilizations increase faster (possibly due to a few congested paths in the topology), thus saturating the network faster. Hence $R$ is placed in the denominator.

Also note that we have used Power instead of Energy/bit. The reasoning is one of limitations in estimation, explained as follows. The total energy is the sum of static energy dissipation and dynamic energy. Static energy depends on the execution time, as it is the product of static power and time. In the design-space exploration phase, it is not possible to estimate the improvements in total execution time, because applications may behave differently toward different speed networks. For instance, an application with minimal inter-processor communication may not request more traffic even if the network latency is very low or throughput is high. However, knowing the injection rate, it is possible to estimate the total power during the course of execution of the application. The estimation technique is outlined in the text that follows.

*NoC Parameters:* For the networks in Fig. 2a and Fig. 2b, we adopt the parameters as listed in Table II. The base inter-core spacing is taken as 1 mm. As noted earlier, we are limited to 50 Gb/s links, as HyPPI supports a single wavelength. With a flit size of 64 bits, we needed to make sure that the electronic links match this rate, and thus use 0.78125 GHz. If we did not use equal data rates for all links, there will be extra buffering needed at the transmitter/receiver ends of the optical links; this is a separate issue in network design, which is not the focus of this work. With 50 Gb/s links, photonics needs two wavelengths. We could conceivably use a larger number of wavelength for photonics, but that will need more MRRs (rings), thereby increasing the thermal trimming power. More MRRs also contribute to losses on the waveguide, which would demand higher laser power. For energy and area estimations, we used the DSENT tool [8] for an accurate analysis, using 11 nm technology node. For HyPPI, we modified DSENT based on the component parameters listed in Table I (except, of course, the energy efficiency numbers listed in the table, because power is computed by DSENT; photonic modulator insertion loss and extinction ratio are also optimized by DSENT).

Table II
NETWORK PARAMETERS USED FOR ALL NOCS IN THIS WORK

| Parameters | Value |
|---|---|
| **Nodes** | |
| # Nodes | 16×16 (256 nodes) |
| Core Spacing | 1 mm |
| Core Clk Freq | 0.78125 GHz |
| **Router** | |
| Flit Size | 64 bits |
| # Ports | 5 (base) or 7 (hybrid) |
| # Virtual Channels (VCs) | 4 |
| Buffers per VC | 8 flits |
| Pipeline Depth | 3 stages |
| Area | From Modified-DSENT |
| Static Power | From Modified-DSENT |
| Dynamic Energy per Flit | From Modified-DSENT |
| **Link** | |
| Latency | 1 clk Elec.,else 2 clks |
| Capacity | 50 Gb/s |
| Area | From Modified-DSENT |
| Static Power | From Modified-DSENT |
| Dynamic Energy per Flit | From Modified-DSENT |

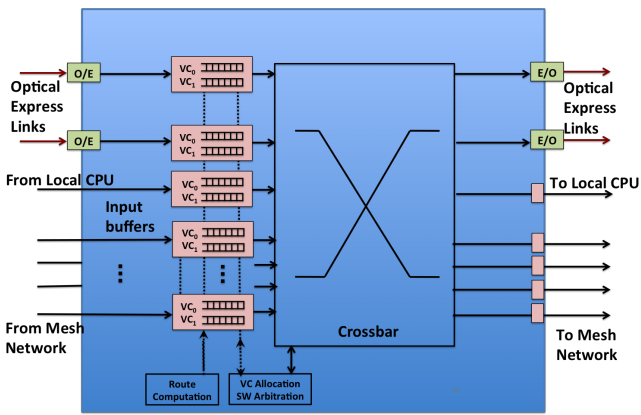

Figure 4. Hybrid NoC Router with Optical Express Links.

The latency value for electronic links is 1 clock cycle. The architecture of the hybrid router, with base electronics augmented by optical express links, is shown in Fig. 4. Only two virtual channels are shown for simplicity; router parameters are listed in Table II. As depicted, the basic routing always uses electronics. Since express links are bidirectional, the hybrid router needs two additional ports. For optical express links, an additional clock cycle is required for O-E conversion at the receiver. At the sending end, the router output buffer already has a register which is utilized as a staging area for the transmitter, and thus the transmitter pipeline is already accounted for within the router delay. This is an accurate assumption because each link has a dedicated router port. Each E-O conversion implies an on-chip laser. The link propagation delay, for all lengths that we consider, is bounded within one clock period (i.e. 1 clock @ 0.78125 GHz). Thus the total latency for all optical links is 2 clock cycles.

*Evaluation Methodology*: With the different networks as set up in Fig. 2a and Fig. 2b, we then used synthetic traffic statistics to model input traffic, based on Soteriou et. al [15], with $p = 0.02$ $\sigma = 0.4$. The value of $p$ signifies the acceptance probability of a flit, and thus captures the spatial hop distribution. Low $p$ implies longer hops. The value of $\sigma$ gives the standard deviation of the traffic injection which follows a gaussian distribution; a larger value implies more nodes are injecting traffic. The maximum injection rate is set to 0.1. This value is typical for NoCs, and capturing the benefits of optical networks at these injection rates is crucial [8] since optical links have high static power. Optical links (such as nanophotonics and HyPPI) typically show good performance at high injection rates, since the static power is amortized across their high data rate. Hence realistic injection ratios are important. For the traffic model, we used only flit counts between source-destination pairs, and temporal information is ignored (except for the injection rate).

After setting up the traffic, each network was then analyzed in order to compute the resulting injection rate across every link in the network. An oblivious shortest-path routing method was adopted, in order to match the routing technique used in the BookSim 2.0 simulator [16] for custom networks, the simulator we use in a later section in this paper.

Based on the injection rate information obtained for each link, the power consumption was computed based on the static power and dynamic energy per flit numbers from DSENT. This was carried out across all network components, the links and routers.

Based on the resulting estimate for the average utilization of the network, R was also calculated according to equation (3). The average latency is also estimated based on the shortest paths, using the individual latency values for the links and routers.

*Base Mesh*: Based on the outlined methodology, we computed CLEAR using equation (2) for the base mesh network created using different technologies. Results are shown in Fig. 5, first bar. The units used were, Gb/s for C, clks for L, Watts for E (power), and $mm^2$ for A. The respective Latency, Power, and Area values are also shown in Fig. 5, in order to understand the CLEAR trends. The

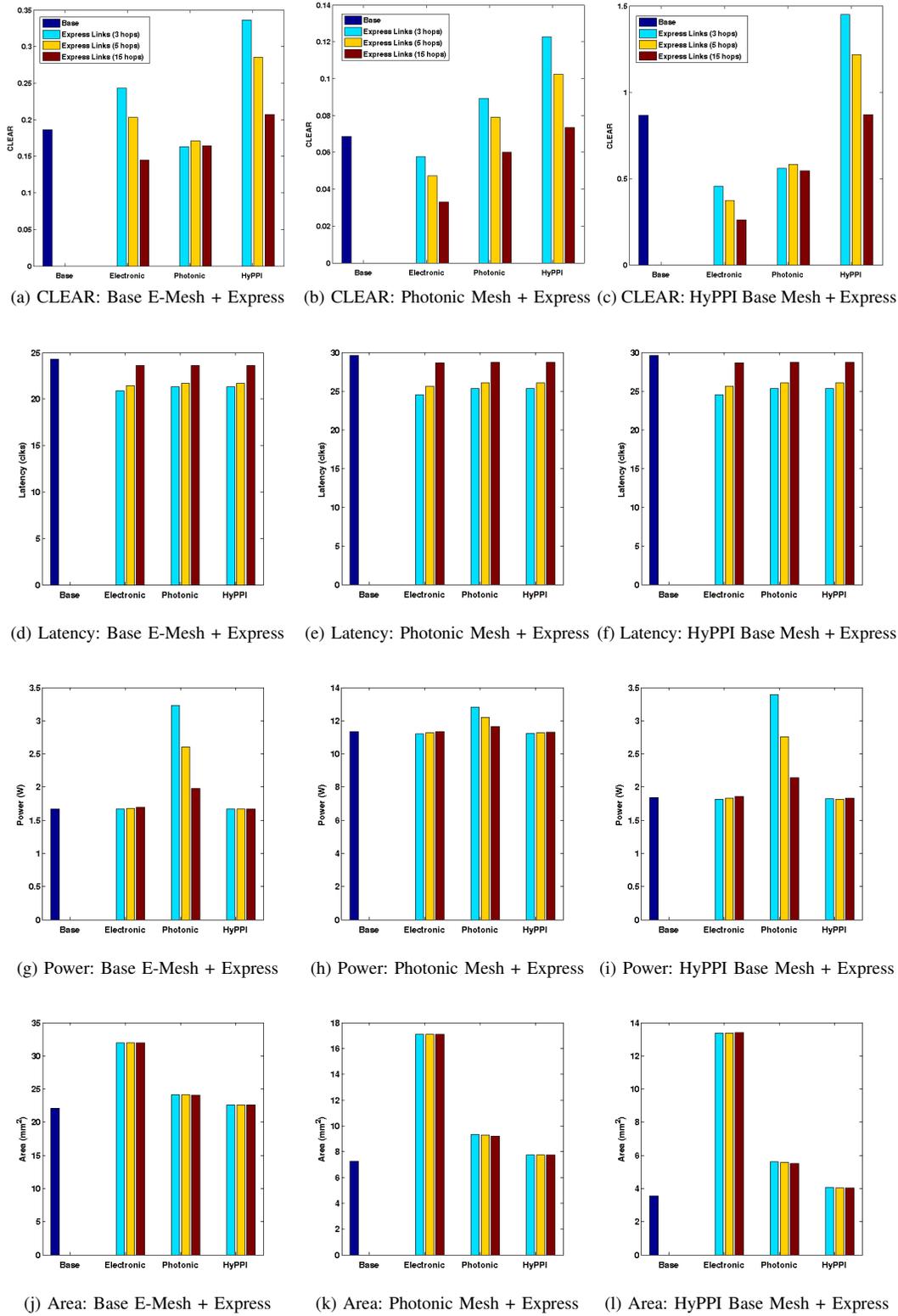

Figure 5. Comparing different flavors of hybrid NoCs (injection rate = 0.1).

Table III
CAPABILITY (C) AND RATE OF UTILIZATION INCREASE (R)

| Parameter | Express Links Topology | | | Plain Mesh |
|---|---|---|---|---|
| | 3 Hops | 5 Hops | 15 Hops | |
| C (Gb/s) | 218.75 | 206.25 | 193.75 | 187.5 |
| R | 0.808 | 0.885 | 1.050 | 1.122 |

Capability (C) and Rate of utilization increase (R) are fixed for a given topology across all technology options, and are summarized in Table III. We observe that for the base mesh network, HyPPI seems to be the best option in terms of CLEAR. Note that the routers are still electronic, for all cases considered. Despite the large number of O-E-O conversions, HyPPI being a plasmonic interconnect has lower energy requirements (comparable to electronics) and also has a much smaller footprint, thus demonstrating a large CLEAR. The latency, however, is poorer for both the optical interconnects (photonics and HyPPI), due to the O-E-O conversion overheads resulting in an additional clock cycle per hop. Nevertheless, HyPPI fares very well in terms of area, as each waveguide of HyPPI has less than $5\mu m$ width (including the pitch). On the other hand, each electronic wire is 160nm wide with 160nm spacing, and thus a 64-bit link requires around $20\mu m$ in width, thus giving a larger area overhead for the network. Photonics has a larger footprint than HyPPI and high static power overhead, and thus fares poorly. We also varied the injection rate from 0.01 to 0.1, and noticed only a small reduction in CLEAR value with the injection rate, and hence we did not plot those here.

*Hybrid NoCs:* We explored the various hybrid NoCs using different technologies for the base Mesh network. With a fixed base network, we tried out Express Links using all four technology options, with different number of hops. Note that the base network Mesh can use any technology. We observe some interesting trends here. Consider the base electronic mesh augmented with express links of different technology types, outputs shown in Fig. 5a,5d,5g, and 5j. Augmenting with photonics long links is the worst option in terms of CLEAR, poorer than electronic long links. We can attribute this to a significant increase in power due to the photonics links. On the other hand, a reverse trend is observed when we adopt photonics as the base mesh network, Fig. 5b,5e,5h, and 5k. Here, using photonics for long links only improves CLEAR, compared with adding electronic long links. This is due to the fact that the base photonic mesh already expends a lot of power, hence the added long links do not considerably increase the power; but they add benefits in terms of reduced area compared with long link electronics. Photonics uses only one waveguide per link, and thus needs less space. In all the plots, we notice that increasing the hop length reduces CLEAR. This is because there are fewer long links incorporated when the number of hops is higher (thus, capacity C is lower, and

Table IV
STATIC POWER, WITH ELECTRONIC BASE MESH + EXPRESS LINKS OF DIFFERENT TECHNOLOGIES

| Express Links Technology | Total NoC Static Power (W) | | |
|---|---|---|---|
| | 3 Hops | 5 Hops | 15 Hops |
| Electronic | 1.532 | 1.533 | 1.547 |
| Photonic | 3.076 | 2.458 | 1.839 |
| HyPPI | 1.545 | 1.539 | 1.533 |
| Static Power for Base Electronic Mesh: 1.53 W | | | |

R is higher, see Table III). For instance, in a $16\times16$ NoC, with Hops=3 we have 5 waveguides per direction in each row (see Figure 2b); whereas with Hops=5, we have only 3 waveguides per direction in each row. We need waveguides for each direction to ensure that the links are bidirectional.

In all cases, we note that HyPPI as the base mesh network provides the best results in terms of CLEAR value. However, if the lowest latency is the target, then a base electronic mesh is the better option, augmented with HyPPI links to minimize area and power overheads. In terms of total power, the base HyPPI and base electronic options are similar. Area-wise, the base HyPPI mesh with augmented HyPPI links gives the lowest overhead.

The final choice of hybridization depends on the specific requirements. For further analysis, we use an Electronic base mesh keeping performance (low latency) as the target. As we can see in Figure 5, augmenting an electronic mesh with HyPPI can give a CLEAR improvement by up to $1.8\times$ (for Express Hops = 3).

## IV. EXPERIMENTAL RESULTS

In order to evaluate the hybrid NoCs further, we use Electronics as the base technology, and augment it with express links. For NoC simulations, we use traces from benchmark suites that run on parallel HPC platforms. We use BookSim 2.0 simulator [16] in trace mode to obtain latency estimates. The network parameters are as listed in Table II. For energy estimates, we obtain the dynamic energy consumption per flit from our modified DSENT, and use it to compute the total dynamic energy based on the communication volume and the network paths taken by the flits. Static power consumption for the different networks is summarized in Table IV. All traffic traces are based on 256-node benchmarks, as the network has a $16\times16$ configuration.

We used the NAS Parallel Benchmarks (NPB) [17], Class A workloads. The following kernels were used - FT, CG, MG, and LU. These benchmarks were executed on a Cray XE6m supercomputer and traffic traces obtained using MPICL. The traces were then converted into BookSim-compatible traces. At times, the traces contained packets that were of very large size (hundreds of kilobytes). Such packets were broken down into smaller packets. For simplicity, all simulations used two types of packets - 1 flit per packet and 32 flits per packet. All large packets from the original network trace were split up into smaller packets, and injected into the network at their respective source nodes, respecting the link bandwidths. The inter-node and intra-node bandwidth in the Cray is in the range 10-13 GB/s, which exceeds our link bandwidth (50 Gb/s, or 6 GB/s), so the traces will not saturate the NoC simulator.

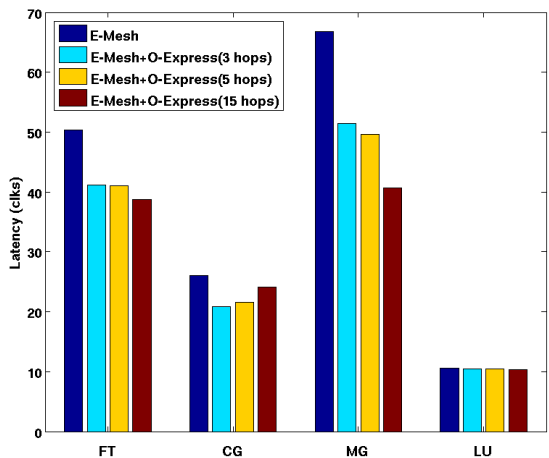

Figure 6. Average Latency for NAS Parallel Benchmarks. The optical technology in O-Mesh could be either Photonics or HyPPI.

Table V
DYNAMIC ENERGY, FT BENCHMARK

| Express Links | Total Dynamic Energy (J) | | |
|---|---|---|---|
| Technology | 3 Hops | 5 Hops | 15 Hops |
| Electronic | 0.0054 | 0.0066 | 0.0128 |
| Photonic | 0.9353 | 0.9353 | 0.9353 |
| HyPPI | 0.0049 | 0.0049 | 0.0049 |
| Dynamic Energy for Plain (Base) Electronic Mesh: 0.0042 J | | | |

*A. Latency*

The latency results for the networks are summarized in Fig. 6. Latency in clock cycles is shown for base Electronic Mesh, as well as Express Links NoCs with Hops = 3, 5, and 15. Hops=15 makes the network effectively a 2D torus. The optical express links could use either photonics or HyPPI. The latency is the same in both cases, because their individual link latencies are identical (2 clks) as previously noted. Thus we don't differentiate between them in Fig. 6. As expected, adding express links reduces the average latency. The CG benchmark has short range traffic and is thus benefited maximum by Hops=3, showing a latency reduction by a factor of 1.25×. On the other hand, MG has long range traffic and thus benefits from longer hops, exhibiting a 1.64× latency reduction for Hops=15. The LU benchmark is almost completely comprised of 1-hop traffic, and thus doesn't derive significant latency improvements. On the other hand, FT has all-to-all traffic, and thus benefits from all types of express links, with a maximum of 1.3× using Hops=15.

*B. Energy*

Augmenting the electronic mesh thus brings notable benefits to latency for real applications. To check whether these improvements come at the cost of energy, we look at the results for the FT benchmark in the NPB suite, Table V. Compared to the base mesh, the hybrid NoC with HyPPI-based express links hardly shows any increase in the relative dynamic energy. In addition, the static power of the electronic mesh is 1.53 W, which is not significantly lower than the 1.533-1.545 W for the hybrid NoC with HyPPI express links, see Table IV. The trends in dynamic energy for the other benchmarks of the NPB suite are very much similar, and are thus not reported here.

Adopting an electronic base mesh with HyPPI express links is therefore an excellent option for building hybrid NoCs. It is still a hybrid network, and incurs overheads for O-E-O conversion at every hop. Fully optical NoCs could potentially provide higher performance, energy, and area benefits, and we explore this further in the next section.

## V. PROJECTIONS FOR ALL-OPTICAL NOCS

Fully optical NoCs, comprised of optical routers and links, are a promising option that need to be evaluated further. Specifically, designers can better leverage the advantages of photonics by using its routing capabilities based on multiple wavelengths, or WDM. There have been all optical NoCs proposed in the literature [1]; however, all of them use a variety of values for the optical parameters making comparisons difficult, as described in our prior work [18]. So we instead construct our own all-optical NoC, Fig. 2c, and use a uniform set of parameters. All-optical NoCs are fundamentally circuit-switched, which means that a path between a source and destination needs to be first established before initiating bulk transfer of packets. Once the path is set up, the latency is one clock cycle or few clock cycles, depending on the path length. However, the exact latency savings is application dependent. Moreover, photonics with ring resonators are very bulky as noted in Section I, and photonic routers typically deploy a large number of rings. Thus, photonic NoCs may not be a very attractive option. All-photonic NoCs may also have other disadvantages, as noted in the beginning of Section III.

However, HyPPI, with its tiny device footprint, can be advantageous for all-optical NoCs. In order evaluate an all-HyPPI NoC, we need a router. We designed a router based on prior art [19]. The building block is our ultra-compact plasmonic electro-optical 2 × 2 switch [20]. The device operates by tuning the coupling length between two SOI waveguide busses by changing the effective index of a

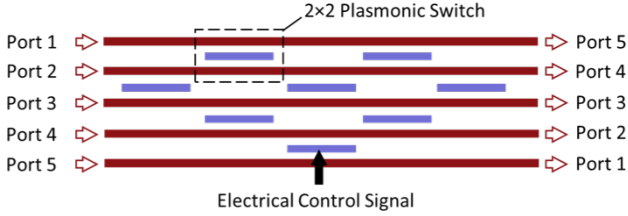

Figure 7. HyPPI Router.

MOS plasmonic island electrically. Due to the compact size ($< 5\mu m$) this switch has fJ/bit power consumption and ps switching delay times. The designed router is depicted in Figure 7, summarized in Table VI. The corresponding photonic router characteristics, which uses 8 rings to realize the eight 2×2 switches [21], are also summarized in the table. The loss incurred by light propagating through the router depends on the input and output port selected. Although the HyPPI router shows potentially larger losses, we are able to use an optimal port assignment (mapping of router ports to the NoC node ports) to incur minimal losses, assuming X-Y dimension ordered routing strategy[†]. Further details of this router are beyond the scope of this paper.

Table VI
COMPARISON OF WDM-BASED PHOTONIC AND HYPPI ROUTERS

| Technology | Control Energy (fJ/bit) | Loss Range (dB) | Area ($\mu m^2$) |
|---|---|---|---|
| Photonic | 68.2 | 0.39-1.5 | 480,000 |
| HyPPI | 3.73 | 0.32-9.1 | 500 |

We then estimated the energy consumption for 16×16 all-optical NoCs for the synthetic traffic used in Section III-B. In order to do so, the losses incurred along the entire path from source to destination for each flit was computed, and the laser power was estimated accordingly using HyPPI and photonic energy equations [9]. Latency values are more challenging to estimate, since the optical path setup will incur overheads. Nevertheless, previously published results reported around 50% reduction in latency over an electronic mesh, with an all-optical NoC using an electronic control network for path setup [22]. We adopt this approximation in our estimates. The area is estimated by using the optical router parameters from Table VI, and the electronic router parameters from the DSENT tool [8].

A comparison of the three networks - electronic mesh, all-photonic NoC, and all-HyPPI NoC - is shown through a radar plot in Fig. 8. Since all the three parameters, namely, Latency, Energy/bit, and Area, are cost values, they need to be small. Thus a triangle with a smaller enclosed area is the better option. From the figure, we note that all-HyPPI has the potential to significantly outperform the other two NoCs. In terms of energy, all-photonic and all-HyPPI are close, at 352 fJ/bit and 354 fJ/bit respectively, whereas electronic mesh requires 89.7 nJ/bit. Area-wise, all-HyPPI fares very well with 1.24 mm$^2$. Electronic mesh needs 22.1 mm$^2$, while the all-photonic mesh needs 127.7 mm$^2$. Thus, if all-optical NoCs are to be considered in the future, HyPPI is an excellent candidate for serving as the technology of choice.

[†]U-turns are not implemented, e.g. connection from Port 1 to Port 1.

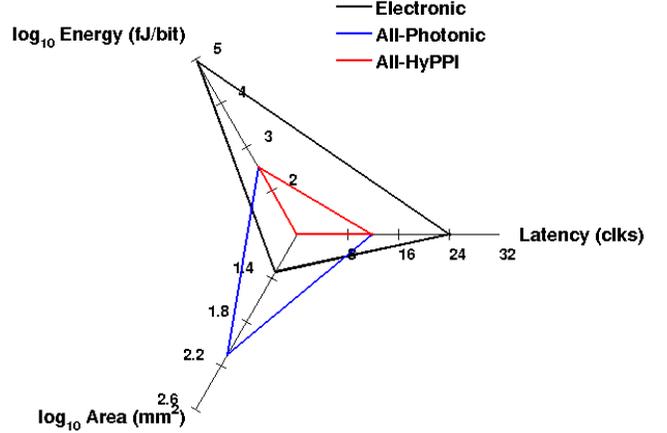

Figure 8. All-Optical WDM NoC vs. Electronic NoC. The triangle that encloses smaller area is the better option.

## VI. CONCLUSIONS

In this paper, we explored new optical link technology options for augmenting on-chip network topologies, namely, hybrid plasmonics - an optical interconnect that uses plasmonics for active data manipulation and diffraction-limited photonics for passive propagation (HyPPI). Next, in order to help design hybrid networks incorporating express links, we adopted a unified metric called CLEAR, and demonstrated results for link and network evaluations using this metric. These evaluations demonstrated that electronic NoCs augmented with HyPPI provided a 1.8× improvement in CLEAR over a base electronic mesh. In carrying out our evaluations, we modified the DSENT tool for modeling HyPPI, in order to obtain accurate energy and area estimates. Then, network level simulations were carried out on the BookSim 2.0 simulator, based on traffic traces from the NAS Parallel Benchmark suite. These results indicated up to 1.64× latency improvement over a base electronic mesh, with negligible energy overheads due to the HyPPI express links. Finally, we carried out performance projections for all-optical NoCs. The projections indicate that all-HyPPI as well as all-photonic NoCs would be significantly more energy efficient than electronic NoCs (255×), although electronic route setup requirements may diminish this result. Furthermore, an all-HyPPI NoC would be two orders of magnitude smaller in area compared with an all-photonic

NoC, and one order of magnitude smaller than an electronic NoC. Thus, HyPPI was demonstrated to be an excellent technology choice for the future, for both hybrid NoCs as well as all-optical NoCs.

## VII. ACKNOWLEDGMENTS

This work was partially supported by the Air Force Office of Scientific Research (AFOSR) under Award FA9550-15-1-0447.